\documentclass[twoside,11pt,letterpaper]{article}

\usepackage{titlesec}
\usepackage{latexsym}
\usepackage{theorem}
\usepackage{fancyhdr}
\usepackage{amssymb}
\usepackage{amsmath}

\renewenvironment{abstract}[1]{\begin{minipage}[t]{\textwidth}
\mbox{\Large #1} \hspace{-0.5em}}{\end{minipage}}

\newcommand{\nn}{\nonumber}

\newcommand{\beq}{\begin{equation}}
\newcommand{\eeq}{\end{equation}}
\newcommand{\beqarr}{\begin{eqnarray}\vspace{1em}}
\newcommand{\eeqarr}{\vspace{1em}\end{eqnarray}}

\newcommand{\bdefinition}{\begin{definition}}
\newcommand{\edefinition}{\end{definition}} 
\newcommand{\btheo}{\begin{theorem}\vspace{1em}}
\newcommand{\etheo}{\vspace{1em}\end{theorem}}

\newcommand{\laction}{\triangleright}

\renewcommand{\vec}[1]{\textbf{#1}}
\newcommand{\tensor}{\otimes}

\newcommand{\id}{\textrm{id}}
\newcommand{\coproduct}{\Delta}
\newcommand{\counit}{\epsilon}
\newcommand{\one}{\mathbf{1}}
\newcommand{\antipode}{\textrm{S}}

\newcommand{\hopfalgebra}{\mathcal{H}}

\newcommand{\liealgebra}{\mathfrak{g}}
\newcommand{\heisenberg}{\mathfrak{h}}
\newcommand{\functions}{\mathcal{F}}
\newcommand{\field}{\mathbf{K}}

\newcommand{\R}{\mathbb{R}}
\newcommand{\C}{\mathbb{C}}
\newcommand{\N}{\mathbb{N}}
\newcommand{\manifold}{\mathcal{M}}

\newcommand{\twist}{\mathcal{F}}
\newcommand{\leftcrossproduct}{>\!\!\!\triangleleft}
\newcommand{\unit}{\eta}

\newcommand{\product}{\mu}

\theoremstyle{change}

\theoremheaderfont{\scshape}

\newtheorem{definition}{Definition}[section]
\newtheorem{theorem}[definition]{Theorem}

\newtheorem{proposition}[definition]{Proposition}

\titlespacing{\section}{0em}{2em}{2em}
\titlespacing{\subsection}{0em}{2em}{2em}

\titleformat{\section}
  {\sc \large}
  {\thesection}
  {1em}
  {}
\titleformat{\subsection}
  {\sc}
  {\thesubsection}
  {1em}
  {}

\numberwithin{equation}{section}

\begin{document}


\title{
\begin{flushright}
	\small \scshape LMU-ASC 53/06 \\
	\small \scshape MPP-2006-98\\[1.7em]
\end{flushright}
\Large \scshape Twist-Deformed Lorentzian Heisenberg-Algebras}
\author{\renewcommand{\thefootnote}{\arabic{footnote}} \scshape Florian Koch\footnotemark[1] \\[2em]
\rule{20mm}{0.2mm} \\[1em]
\small \scshape Arnold Sommerfeld Center for Theoretical Physics\\[-0.4em]
\small \scshape Ludwig-Maximilians-Universit\"at M\"unchen\\[-0.4em]
\small \scshape Theresienstra{\ss}e 37, 80333 M\"unchen, Germany\\[0.8em] 
\small \scshape Werner Heisenberg Institut\\[-0.4em]
\small \scshape Max-Planck-Institut f\"ur Physik\\[-0.4em]
\small \scshape F\"ohringer Ring 6, 80805 M\"unchen, Germany\\[1em] 
\rule{20mm}{0.2mm} \\[2em]}
\date{}

\footnotetext[1]{koch@theorie.physik.uni-muenchen.de}

\maketitle


\begin{abstract}{T}he Moyal-Weyl quantization procedure is embedded into the twist formalism of vector fields 
on phase space. Double application of twists provide most general deformations of Minkowski\-an Hei\-sen\-berg-al\-ge\-bras and corresponding quantizations of the Lorentz-algebra. Such 
deformations deliver high-energy extensions of standard relativistic quantum mechanics. These are
required to obtain minimal uncertainty properties for high-energy spacetime measurements that
standard quantum mechanics lacks. The procedure of double twist application is outlined. We give 
an instructive and genuine example.
\end{abstract}

\setcounter{footnote}{1}
\thispagestyle{empty}



\newpage

\section{Introduction}

The scheme of canonical quantization, presented in textbooks of quantum mechanics, is the most 
simple quantization one might perform. Noncommutative geometry is considered as some enhancement 
of this scheme. There are two basic ideas of how noncommutative geometry can be 
interpreted in physics. From the side of effective theories, we hope for some alternatives
to standard perturbative treatment of field theories and their renormalization. Such alternatives
would be required by quantum chromodynamics and gravity such as \cite{Aschieri:2005yw,Aschieri:2005zs} 
already suggests. On the other hand one might stick to a more fundamental point of view. Noncommutative 
geometry is then regarded as a gravity effect itself. Such approaches can be found in gravity 
motivated canonical noncommutative geometry \cite{Doplicher:1994tu,Doplicher:1994zv}, but also within 
discussions of minimal uncertainty theories such as in \cite{Kempf:1998em,Kempf:1998gk,Kempf:1997sr,
Kempf:1994qp}. Moreover there are close relations of noncommutative geometry as well as of 
doubly special relativity to loop quantum gravity \cite{Amelino-Camelia:2004ht, Amelino-Camelia:2003uc, Amelino-Camelia:2005et,Amelino-Camelia:2003xp}. Within such a fundamental approach, noncommutative 
geometry should not be expected as a static noncommutative background for field theories anymore. Instead, noncommutative geometry itself should become subject to gravity by making it dependent on energy and 
momentum. After all we expect, Planck scale effects at high energy-momentum densities and thus a grainy 
structure of spacetime, obtained from noncommutative geometry, can only be mediated by operators 
of energy and momentum. This is nothing else than a more general deformation of phase space than 
obtained by canonical quantization. Moreover in such an approach, noncommutative geometry should become 
localized to those space volumes, where densities of energy and momentum enter the actual high energy 
regime. Standard problems such as IR-UV-Mixing effects should thus not occure in such a setup.
A first and actually most prominent example of such a \emph{general quantization} is the well known 
Quantum-Spacetime of Snyder \cite{Golfand63b,Golfand63a,Golfand60,Snyder:1946qz,Snyder46b,Yang:1947ud}. 
Canonical quantization can be understood as a deformation-quantization of the phase space towards the Heisenberg-algebra. Weyl and Moyal \cite{Moyal:1949sk,Weyl:1927vd} performed this deformation by means of starproducts. In this paper we formalise this setup by introducing a Hopf-algebra of vector fields on 
phase space. We use these vector fields to twist the phase space to the standard Heisenberg-algebra. 
In a second step we further apply twists to deform the Heisenberg-algebra itself. These two twists can 
be merged to a single one. The paper is organized as follows. In the first section we introduce the 
$2n$-dimensional Heisenberg-algebra $\heisenberg_{2n}$ and its universal enveloping algebra 
$U(\heisenberg_{2n})$. We then recall how this algebra is obtained by deformation-quantization of a 
commutative phase space algebra. This is due to Weyl and Moyal. We formalise and introduce a 
Hopf-algebra of vector fields on the phase space. In the following we discuss twisting by means of 
these vector fields. To this purpose we show that the product of two twist once more is a twist. 
We further present basic examples and discuss results in a conlusion. 

Before we actually come to general matters, we first have to do some remarks that clearify and motivate 
the directions pursued in the following constructions and that indeed go hand in hand with the formalism 
chosen by Weyl and Moyal. 

In textbooks on field theory, we often find the representation of the Lorentz-algebra in terms of generators of $U(\heisenberg_{2n})$. In particular the generators $m^{\mu\nu}$ of the Lorentz-algebra are represented in $U(\heisenberg_{2n})$ by
$$
	m^{\mu\nu} = x^\mu p^\nu - x^\nu p^\mu.
$$
Using the commutation relation
\beq
	\left[p^\mu, x^\rho \right] = - i \eta^{\mu\rho}, \label{xpaction}
\eeq
the \emph{action} of $m^{\mu\nu}$ on basis elements $x^\rho$ and $p^\sigma$ of $U(\heisenberg_{2n})$ is then 
evaluated by commutators
\beqarr
	\left[m^{\mu\nu}, x^\rho \right] & = & \left[x^\mu p^\nu - x^\nu p^\mu, x^\rho \right]
	= x^\mu \left[p^\nu, x^\rho\right] - x^\nu \left[p^\mu, x^\rho\right] \nn \\
	& = & -i \eta^{\nu\rho} x^\mu + i \eta^{\mu\rho} x^\nu \label{xmaction} \\
	\left[m^{\mu\nu}, p^\sigma \right] & = & \left[x^\mu p^\nu - x^\nu p^\mu, p^\sigma \right]
	= \left[x^\mu, p^\sigma \right] p^\nu - \left[x^\nu, p^\sigma \right] p^\mu \nn \\
	& = & i \eta^{\mu\sigma} p^\nu - i \eta^{\nu\sigma} p^\mu. \label{pmaction}
\eeqarr
There are several pictures how this setup can be interpreted in physics. At first we can stick to the 
Poincar\'e-algebra, generated by $m^{\mu\nu}$ and $p^\rho$, that is represented on Minkowski-space.
In this scheme we do not consider the Lorentz-algebra to be represented in terms of generators of
$U(\heisenberg_{2n})$, as we did above - but nevertheless consider the "representation" of the 
Lorentz-algebra in terms of commutators $\left[m^{\mu\nu}, x^\rho\right]$ or 
$\left[p^\nu, x^\rho\right]$ althought this already incorporates a multiplicative structure 
between the symmetry algebra and its representation space. For the commutative case this is 
alright - but deformations to noncommutative geometry modify the commutation relations in such a 
way that they do not close on the representation space anymore. There is actually a mixing of the 
symmetry algebra and the representation space. This phenomenon is also discribed in 
\cite{Zachos:2001ux}. To fix this problem we might thus argue that we have actually to stay 
within the Heisenberg-algebra $U(\heisenberg_{2n})$. Then, with $m^{\mu\nu}$ represented in 
$U(\heisenberg_{2n})$ as performed above, we do not care anymore if a mixing occures. In this 
case the commutator $\left[p^\nu, x^\rho\right]$ manages everything that is represented on 
Minkowski-space. At first this argumentation makes perfect sense and in the case of deformations
of Minkowskian $U(\heisenberg_{2n})$ it has been reasoned a long such a way \cite{Kempf:1998em,Kempf:1998gk,Kempf:1997sr,Kempf:1994qp}. Algebraically the subalgebra of momenta 
in $U(\heisenberg_{2n})$ does not differ from that of coordinates and thus if the commutator 
$\left[p^\nu, x^\rho\right]$ is considered to represent the subalgebra of momenta on the 
coordinates, we might as well argue that in turn $\left[x^\mu, p^\sigma \right]$ is some sort 
of representation of coordinates on the momenta as also performed in our computation in 
(\ref{pmaction}) from above. But this as well rises the question how a coordinate would 
possibly act on products of generators of momenta. Or in other words, what is the coproduct 
of a coordinate ? This argumentation is of course too naive and these issues actually do not become 
a question for the commutative case - but if we are to consider deformations, we have to know about 
such coproducts, at least in principle. We have to have a neat bialgebra or Hopf-algebra 
as a framework to consider any deformation. In fact it is not possible to endow the coordinates 
with the same \emph{primitive type} of coproduct as we use it for the momenta. Such an 
introduction of a coproduct contradicts the property of the coproduct to be 
an algebra-homomorphism. Nevertheless there are examples that neatly and quite elegantly endow
a phase space with proper coproducts on momenta and coordinates \cite{Majid:1998up}. 
However these also incorporate some specific structure that already accomodates some physics.
The solution to this dilemma can be found in the introduction of vector fields on the entire 
phase space that we are presenting here. This had been performed first by Moyal and Weyl in \cite{Moyal:1949sk,Weyl:1927vd}. We thus first concentrate on their work in a Minkowkian setting 
and formalize this to our requirements. In particular we lift these vector fields to a 
Hopf-algebra as presented in \cite{Koch:2006cq}. We are then able to fit in the Lorentz-symmetry 
and consider further deformations.

\section{Quantum Mechanics according to Weyl and Moyal}

This section is intended as a basic review and outline that constitutes the actual input and fundaments of 
our constructions. The section is divided in two parts. In the first subsection we introduce $n$-dimensional
Minkowski-space and the corresponding representation of the Poincar\'e-algebra. This is the only input we 
require for all of our considerations in this work. Based on this we build the $2n$-dimensional Minkowskian 
phase space and the Heisenberg Lie-algebra by taking direct sums of copies of Minkowski-space. These three 
vector spaces are further more enhanced to algebras of universal enveloping algebra type. The second subsection
then reviews the deformation-quantization of Minkowskian phase space towards the Heisenberg-algebra according 
to Weyl and Moyal using the starproduct. In mathematical terms this is a deformation-quantization of a Poisson-Manifold. For completeness we shortly review this latter notion. We thereby obtain the required 
setup for further deformations with the double application of twists that is discussed in the next sections. 
As a textbook we recommend \cite{Charipressley} as reference for this section.

\subsection{The Minkowskian Heisenberg-Algebra}

The $n$-dimensional Minkowski-space $\R^{(1, n-1)}$ is a vector space with scalar product
\beq
	<\vec{x}, \vec{y}> = \eta_{\mu\nu} \; x^\mu y^\nu, \;\;\;\; \vec{x}, \vec{y} \in \R^{(1, n-1)},
	\label{minkowskiproduct}
\eeq
that is left invariant under the action of the Lorentz-group $\textrm{SO}(1,n-1)$. Within a specific 
coordinate system, the invariance of (\ref{minkowskiproduct}) under matrix representations of transformations 
$\Lambda \in \textrm{SO}(1,n-1)$ is given by
$$
	\eta_{\rho\sigma} = \eta_{\mu\nu} \; \Lambda^\mu_{\;\;\;\; \rho} \; \Lambda^\nu_{\;\;\;\; \sigma},
	\;\;\;\; \mu, \nu, \rho, \sigma \in 0, \ldots (n - 1).
$$
The signature of the metric tensor $\eta_{\mu\nu}$ has not to be specified within our consideration. 
We consider Minkowski space to be generated by a basis $(x^\mu)_{\mu \in 0, 1, \ldots n-1}$. 
Apart from isotropy of spacetime, homogeneity of $\R^{(1, n-1)}$ is generated by the action of the 
$n$-dimensional translational group $T_n$. The Poincar\'e group $\mathfrak{P}$ is the semi-direct product $\textrm{SO}(1,n-1) \leftcrossproduct \; T_n$. The Lie groups $\textrm{SO}(1,n-1)$ and $T_n$ are generated 
by Lie-algebras $\textrm{so}_{1,n-1}$ and $t_n$ respectively to constitute the Poincar\'e-algebra 
$\mathfrak{p}$. In particular for representations we actually consider the universal enveloping algebras $U(\mathfrak{p})$, $U(\textrm{so}_{1,n-1})$ and $U(t_n)$. In order to endow Minkowski-space with a 
commutative algebraic structure, we enhance it to a Lie-algebra by the introduction of a trivial bracket 
$$
	[\;, \;] \; : \; \R^{(1, n-1)} \times \R^{(1, n-1)} \longrightarrow \R^{(1, n-1)},
$$
that for $x^\rho, x^\sigma \in \R^{(1, n-1)}$ is given by
\beq
	[x^\rho, x^\sigma] = 0. \label{minkowskibracket}
\eeq
On this basis we consider the universal enveloping algebra $U(\R^{(1, n-1)})$. The generators 
$m^{\mu\nu} \in U(\textrm{so}_{1,n-1})$ and $\pi^\rho \in U(t_n)$ of the Poincar\'e-algebra 
$U(\mathfrak{p})$ are subject to commutation relations
\beqarr
	\left[m^{\mu\nu}, m^{\rho\sigma} \right] & = &
		i \eta^{\mu\rho} m^{\nu\sigma} - i \eta^{\nu\rho} m^{\mu\sigma}
		+ i \eta^{\nu\sigma} m^{\mu\rho} - i \eta^{\mu\sigma} m^{\nu\rho}, \nn \\
	\left[m^{\mu\nu}, \pi^\rho \right] & = & i \eta^{\mu\rho} \pi^\nu - i \eta^{\nu\rho} \pi^\mu, \nn \\
	\left[\pi^\rho, \pi^\sigma \right] & = & 0.
	\label{poincarerelations}
\eeqarr
that generate its two-sided ideal. The Poincar\'e-algebra $U(\mathfrak{p})$ becomes a Hopf-algebra
with the following coproduct, counit and antipode:
\beqarr
	& & \coproduct(m^{\mu\nu}) = m^{\mu\nu} \tensor \one + \one \tensor m^{\mu\nu},
		\;\;\;\; \counit(m^{\mu\nu}) = 0,
		\;\;\;\; \antipode(m^{\mu\nu}) = - m^{\mu\nu}, \nn \\
	& & \coproduct(\pi^\rho) = \pi^\rho \tensor \one + \one \tensor \pi^\rho,
		\;\;\;\; \counit(\pi^\rho) = 0,
		\;\;\;\; \antipode(\pi^\rho) = - \pi^\rho.
		\label{coproductpoincare}
\eeqarr
The Hopf-algebra $U(\mathfrak{p})$ is represented on $U(\R^{(1, n-1)})$ as a \emph{left action} by
\beqarr
	m^{\mu\nu} \laction x^\rho & = & -i \eta^{\nu\rho} x^\mu + i \eta^{\mu\rho} x^\nu \nn \\
	\pi^\mu \laction x^\rho & = & - i \eta^{\mu\rho}, \nn \\
	\one_\mathfrak{p} \laction x^\rho & = & x^\rho, \label{reppoincareminkowski}
\eeqarr
such that relations (\ref{poincarerelations}) are realized on the vector space $\R^{(1, n-1)}$, i.e. 
\beqarr
	& & \left( m^{\mu\nu} m^{\rho\sigma} - m^{\rho\sigma} m^{\mu\nu}
		- i \eta^{\mu\rho} m^{\nu\sigma} + i \eta^{\nu\rho} m^{\mu\sigma}
		- i \eta^{\nu\sigma} m^{\mu\rho} + i \eta^{\mu\sigma} m^{\nu\rho} \right) \laction x^\lambda = 0, \nn \\
	& & \left( m^{\mu\nu} \pi^\rho - \pi^\rho m^{\mu\nu}
		- i \eta^{\mu\rho} \pi^\nu + i \eta^{\nu\rho} \pi^\mu \right) \laction x^\lambda = 0, \nn \\
	& & \left( \pi^\rho \pi^\sigma - \pi^\sigma \pi^\rho \right) \laction x^\lambda = 0.
	\label{realizedpoincareminkowski}
\eeqarr
The action of generators $m^{\mu\nu}, \pi^\mu \in U(\mathfrak{p})$ on products of coordinates in  
$U(\R^{(1, n-1)})$ is given by 
\beqarr
	m^{\mu\nu} \laction (x^\rho x^\sigma) & = & \coproduct(m^{\mu\nu}) \laction (x^\rho x^\sigma) 
		= (m^{\mu\nu} \laction x^\rho)x^\sigma + x^\rho (m^{\mu\nu} \laction x^\sigma), \nn \\
	\pi^\mu \laction (x^\rho x^\sigma) & = & \coproduct(\pi^\mu) \laction (x^\rho x^\sigma) 
		= (\pi^\mu \laction x^\rho)x^\sigma + x^\rho (\pi^\mu \laction x^\sigma), \nn \\
	m^{\mu\nu} \laction \one & = & \counit(m^{\mu\nu}), \;\;\;\; 
		\pi^\mu \laction \one \; = \; \counit(p^\mu), \label{productaction}
\eeqarr
such that the generating relations (\ref{minkowskibracket}) of $U(\R^{(1, n-1)})$ are respected by their 
action according to
\beqarr
	m^{\mu\nu} \laction (x^\rho x^\sigma - x^\sigma x^\rho - [x^\rho, x^\sigma]) & = & 0, \nn \\
	\pi^\mu \laction (x^\rho x^\sigma - x^\sigma x^\rho - [x^\rho, x^\sigma]) & = & 0. 
	\label{poincaremapscommutative}
\eeqarr
As a next step we introduce Minkowskian phase space $\Gamma$ as the direct sum of two copies of 
Minkowski-space $\R^{(1, n-1)}$, i.e. we obtain 
\beq
	\Gamma = \R^{(1, n-1)} \oplus \R^{(1, n-1)}. \label{minkowskiphasespace}
\eeq
As for Minkowski-space, we enhance $\Gamma$ with a commutative Lie-algebraic structure.
Within a specific coordinate system we thus take $(x^\mu, p^\nu)_{\mu, \nu \in 0, 1, \ldots n-1}$
as a basis and introduce the brackets
\beqarr
	\left[x^\mu, x^\nu \right] & = & 0, \nn \\
	\left[x^\mu, p^\nu \right] & = & 0, \nn \\
	\left[p^\mu, p^\nu \right] & = & 0, \label{phasespaceideal}
\eeqarr
We then obtain the universal enveloping algebra $U(\Gamma)$ by once more taking these brackets as the 
generating relations for the corresponding two-sided ideal of $U(\Gamma)$. Concerning covariance under 
the action of $U(\mathfrak{p})$, we can replace coordinates $x$ by momenta $p$ in conditions (\ref{reppoincareminkowski}) and (\ref{realizedpoincareminkowski}), i.e. on the vector space 
$\Gamma = \R^{(1, n-1)} \oplus \R^{(1, n-1)}$ the Lorentz group $\textrm{SO}(1,n-1)$ is represented 
by block-diagonal matrices
\beq
\Lambda_\mathcal{P} = \left( \begin{array}{cc}
															\Lambda & 0 \\
															0	& \Lambda 
														\end{array}
 											\right).
	\label{blockdiagonal}
\eeq
With respect to the covariance of the algebraic structure of $U(\Gamma)$ we can replace products
of coordinates $x^\rho x^\sigma$ in (\ref{productaction}) and (\ref{productaction}) by products of
coordinates and momenta $x^\rho p^\sigma$ and products of momenta $p^\rho p^\sigma$. We thereby 
obtained a left action of $U(\mathfrak{p})$ on $U(\Gamma)$. 

In a similar manner, as for $\Gamma$, we obtain the Minkowskian Heisenberg-algebra $\heisenberg_{2n}$ 
by taking the direct sum of two copies of $\R^{(1, n-1)}$ and the real numbers, i.e.
\beq
	\heisenberg_{2n} = \R^{(1, n-1)} \oplus \R^{(1, n-1)} \oplus i \R.
	\label{minkowskiheisenberg}
\eeq
This vector space becomes a Lie-algebra by introducing a bracket
$$
	\left[\;, \; \right] \; : \; \heisenberg_{2n} \times \heisenberg_{2n} \longrightarrow \heisenberg_{2n}
$$
that for $\vec{X}_1, \vec{Y}_1, \vec{X}_2, \vec{Y}_2 \in \R^{(1, n-1)}$ and $c_1, c_2 \in \R$ is 
defined by
\beq
	\left[(\vec{X}_1, \vec{Y}_1, c_1), (\vec{X}_2, \vec{Y}_2, c_2)\right] =
			\left(0, 0, i \cdot (<\vec{X}_1, \vec{Y}_2> - <\vec{Y}_1, \vec{X}_2>) \right). 
	\label{heisenbergbracket}
\eeq
Through the scalar product (\ref{minkowskiproduct}) used in this definition we obtain $\heisenberg_{2n}$
to be covariant under the action of $U(\mathfrak{p})$. Besides this, Lorentz-covariance is equally 
intruduced as for the phase space $\Gamma$. By the identification
$$
	X^\mu \equiv \; (e^\mu, 0, 0) \; \in \; \heisenberg_{2n}, \;\;\;\; 
	P_\nu \equiv (0, e_\nu, 0) \; \in \heisenberg_{2n},
$$
we obtain the bracket-relations between coordinates $X^\mu$ and momenta $P^\nu$
\beq
	\left[X^\mu, X^\nu \right] = 0 \; , \;\;\;\;
	\left[X^\mu, P^\nu \right] = i \hbar \eta^{\mu\nu} \; , \;\;\;\;
	\left[P^\mu, P^\nu \right] = 0.
	\label{heisenbergideal}
\eeq
These relations once more generate the two-sided ideal that is required to formulate the universal 
enveloping algebra $U(\heisenberg_{2n})$ of the Heisenberg-algebra $\heisenberg_{2n}$. We are now
prepared to consider deformation-quantization of $U(\Gamma)$ towards $U(\heisenberg_{2n})$ as it 
had been introduced by Moyal. 

\subsection{Phase Space Quantization with Starproducts}

In the last subsection we considered the phase space algebra as the universal enveloping algebra 
$U(\Gamma)$. Dually we have the algebra of complex-valued functions $\functions(\Gamma)$ on 
$\Gamma$. Defining the Poisson-bracket on functions $\functions(\Gamma)$, we turn $\Gamma$ into
a Poisson-manifold. As such we deform it to $U(\heisenberg_{2n})$ according to the quantization 
procedure applied by Moyal \cite{Moyal:1949sk}. This is more generally known as a deformation of 
Poisson manifolds. We recall these notions here. In order to perform this quantization we switch
between the dual pictures of $U(\Gamma)$ and $\functions(\Gamma)$. We begin by introducing the algebra 
of functions $\functions(\Gamma)$ on $\Gamma$.

On the vector space $\Gamma = \R^{(1, n-1)} \oplus \R^{(1, n-1)}$ we consider the subset 
$\functions(\Gamma) \subset C^\infty (\Gamma, \C)$ of smooth complex-valued functions, that we 
endow with a Poisson-bracket
$$
	\{ \; , \; \} \; : \; \functions(\Gamma) \times \functions(\Gamma)
		\longrightarrow \functions(\Gamma),
$$
that in particular is defined for $\omega, \varphi \in \functions(\Gamma)$ by 
\beq
	\{\omega, \varphi\} 
		:= \frac{\partial \omega}{\partial p_\mu} \cdot \frac{\partial \varphi}{\partial x^\mu}
			- \frac{\partial \omega}{\partial x^\mu} \cdot \frac{\partial \varphi}{\partial p_\mu}.
	\label{minkowskipoissonbracket}
\eeq
In addition to this bracket, the vector space of functions $\functions(\Gamma)$ is endowed with 
pointwise multiplication that is induced from the product within the complex numbers, i.e. 
for $\omega, \varphi \in \functions(\Gamma)$ we have
$$
	(\omega \cdot_\functions \varphi)(x^\mu, p_\nu) = \omega(x^\mu, p_\nu) \cdot_\C \varphi(x^\mu, p_\nu)
$$
By the introduction of the Poisson-bracket (\ref{minkowskipoissonbracket}), we turn the vector space 
$\Gamma$ into what is called a Poisson manifold that is more generally defined as follows.

\begin{definition}
	Let $\manifold$ be a $d$-dimensional manifold and $C^\infty (\manifold, \C)$ be the set of complex-valued 
	smooth functions on $\manifold$. Then $\manifold$ is called a \emph{Poisson Manifold}, if there exists a 
	bracket $\{\cdot, \cdot\}$ 
	$$ \{\cdot, \cdot\} : C^\infty (\manifold, \C) \times C^\infty (\manifold, \C) \to C^\infty (\manifold, \C),$$
	such that the following properties hold:
	\beqarr
		\forall \; \omega, \varphi, \psi \in C^\infty (\manifold, \C) 
			& : & \{\varphi, \omega\} = - \{\omega, \varphi\} \nn \\
		& & \{\varphi \cdot \omega, \psi \} = \varphi \cdot \{\omega ,\psi \} + \{ \varphi, \psi \} \cdot \omega \nn \\
		& & \{\{\varphi, \omega \}, \psi \} + \{\{ \omega, \psi \}, \varphi \} + \{\{ \psi, \varphi\}, \omega\} = 0 \nn
	\eeqarr
\end{definition}

We thus have two distinct algebraic structures on $\Gamma$, i.e. on $\functions(\Gamma)$. The original 
problem considered by Weyl and Moyal in \cite{Moyal:1949sk,Weyl:1927vd} had been to grasp the procedure 
of quantization as mathematical term. The procedure of quantization in particular sends the Poisson-bracket 
of $\functions(\Gamma)$ to the commutator of $U(\heisenberg_{2n})$ according to
$$
	\{ \;, \; \} \longrightarrow \frac{i}{\hbar}\left[ \;, \; \right].
$$
This procedure agitates the former algebraic structures of $\Gamma$. It "maps" the commutative algebra 
of functions $\functions(\Gamma)$ to the noncommutative $U(\heisenberg_{2n})$. The solution is to consider
quantization to be the deformation of the product of the algebra of functions $\functions(\Gamma)$ performed
in such a way that the commutator of the deformed algebra of functions corresponds to the structure 
implied by the Poisson-bracket. More generally this is known to be a quantization of a Poisson-manifold that
more precisely is defined as follows.

\begin{definition}
	Let a Poisson manifold $(\manifold, \{ \cdot, \cdot \}, \field)$ over the field $\field$ be given. 
	A \emph{quantization} of $\manifold$ with deformation parameter $h \in \field$ is a manifold $\manifold_h =   
	(\manifold, [\cdot \stackrel{*_h}{,} \cdot ], \field)$, such that to first order in the deformation 
	parameter $h$ the commutator $[\cdot \stackrel{*_h}{,} \cdot ]$ satisfies the following property:
	$$ \forall f_1, f_2 \in \functions(\manifold): \; \frac{[ f_1 \stackrel{*_h}{,} f_2 ]}{h} 
		= \frac{ f_1 *_h f_2 - f_2 *_h f_1 }{h}
		= \{f_1 , f_2 \} \;\;\;  \textrm{mod}(h)$$
\end{definition}

The quantization of the algebra of functions is typically performed in terms of starproducts. To this purpose
it is convenient to consider $U(\Gamma)$ instead of $\functions(\Gamma)$. Since $\functions(\Gamma) \subset C^\infty (\Gamma, \C)$ and $\functions(\Gamma)$ is commutative, this duality merely means that functions $\varphi \in \functions(\Gamma)$ can be represented in terms of formal power series in $U(\Gamma)$ that moreover can be regarded
as power series of a real parameters and thus can converge locally. We thus express functions $\varphi \in \functions(\Gamma)$ as power series
\beqarr
	\varphi(x^\mu, p_\nu) & = & \sum_{\vec{r}, \vec{s}} C_{\vec{r}, \vec{s}}
	\cdot (x^0)^{r_0} \cdot \ldots \cdot (x^{(n-1)})^{r_{(n-1)}} 
	\cdot (p_0)^{s_0} \cdot \ldots \cdot (p_{(n-1)})^{s_{(n-1)}} \nn \\
	& & C_{\vec{r}, \vec{s}} \in \C; \;\;\;\; \vec{r}, \vec{s} \in \N^n_0 \nn .
\eeqarr
With exponential functions 
$$
	e^{i(\eta_\mu x^\mu + \xi^\nu p_\nu)}, \;\;\;\; \eta_\mu, \xi^\nu \in \R^{(1, n-1)}
$$
as a basis for $\functions(\Gamma)$ we can also consider $\varphi \in \functions(\Gamma)$ as a linear 
combination in terms of its Fourier-transformation
\beqarr
	\varphi(x^\mu, p_\nu) 
		& = & \int d^n \eta \; d^n \xi \; \hat{\varphi}(\eta_\mu, \xi^\nu) 
			\; e^{- i(\eta_\mu x^\mu + \xi^\nu p_\nu)} \nn \\
	\hat{\varphi}(\eta_\mu, \xi^\nu)
		& = & \frac{1}{(2 \pi)^{2n}}\int d^n x \; d^n p \; \varphi(x^\mu, p_\nu) 
			\; e^{+ i(\eta_\mu x^\mu + \xi^\nu p_\nu)}. \nn
\eeqarr
Of course also for $X^\mu, P_\nu \in U(\heisenberg_{2n})$ exponential functions
$$
	e^{i(\eta_\mu X^\mu + \xi^\nu P_\nu)}, \;\;\;\; \eta_\mu, \xi^\nu \in \R^{(1, n-1)}
$$
constitute a basis for $U(\heisenberg_{2n})$. In particular these exponentails are group elements
of the corresponding Heisenberg Lie-group. Note that $U(\heisenberg_{2n})$ is dual to a corresponding 
algebra of functions over the Heisenberg Lie-group. The Poincar\'e-Brikhoff-Witt theorem enables us to
map $U(\heisenberg_{2n})$ to $\functions(\Gamma)$ by an isomorphism $W$ of vector spaces. In particular 
this statement reads as follows.

\begin{theorem}
	Let $\liealgebra$ be an n-dimensional Lie-algebra with basis $(g_i)_{i \in \{1 \ldots n \}}$ over the field 	
	$\field$. Furthermore let 
	\beqarr
		\pi: \{1 \ldots n \} \subset \N & \to & \{1 \ldots n \} \nn \\
		k & \mapsto & i_k \nn
	\eeqarr
	be any permutation, then the ordered monomials 
	$$ 
		(g_{i_1})^{m_{i_1}} \ldots (g_{i_k})^{m_{i_k}} \ldots (g_{i_n})^{m_{i_n}} \in U(\liealgebra), \;\; m_{i_k} \in \N 
	$$
	constitute a basis of the universal enveloping algebra $U(\liealgebra)$ of $\liealgebra$ and there exists an 
	isomorphism $W$ of vector spaces
	\beqarr
		W & : & U(\liealgebra) \to U(\R^{n}) \nn \\
		& & (g_{i_1})^{m_{i_1}} \ldots (g_{i_k})^{m_{i_k}} \ldots (g_{i_n})^{m_{i_n}} 
			\mapsto (x_{i_1})^{m_{i_1}} \ldots (x_{i_k})^{m_{i_k}} \ldots (x_{i_n})^{m_{i_n}}. \nn
	\eeqarr
\end{theorem}

Introducing a starproduct on $\functions(\Gamma)$, i.e. performing the quantization of the Poisson-manifold
as described, actually enhances the isomorphism $W$ of vector spaces to an isomorphism of corresponding 
algebras. In particular we therefore consider how basis elments are mapped, i.e. we obtain
\beqarr
	W : U(\heisenberg_{2n}) & \to & \functions(\Gamma) \nn \\
		e^{i(\eta_\mu X^\mu + \xi^\nu P_\nu)} & \mapsto & e^{i(\eta_\mu x^\mu + \xi^\nu p_\nu)}. \nn
\eeqarr
By application of the inverse map $W^{-1}$ we receive for two functions 
$\varphi, \omega \in \functions(\Gamma)$ the corresponding objects within 
$U(\heisenberg_{2n})$. In particular we obtain
\beqarr
	W^{-1}(\varphi)(X^\mu, P_\nu) 
		& = & \int d^n \eta \; d^n \xi \;\; \hat{\varphi}(\eta_\mu, \xi^\nu) 
			\; e^{- i(\eta_\mu X^\mu + \xi^\nu P_\nu)} \nn \\
	W^{-1}(\omega)(X^\mu, P_\nu) 
		& = & \int d^n \eta \; d^n \xi \;\; \hat{\omega}(\eta_\mu, \xi^\nu) 
			\; e^{- i(\eta_\mu X^\mu + \xi^\nu P_\nu)}. \nn
\eeqarr
In order to endow the vector space $\Gamma$ with a deformed multiplication map $*_\hbar$ we require that
\beqarr
	W^{-1}(\varphi *_\hbar \omega)(X^\mu, P_\nu) & := & W^{-1}(\varphi)(X^\mu, P_\nu) 
		\cdot W^{-1}(\omega)(X^\mu, P_\nu) \nn \\
	& = & \int d^n \eta \; d^n \xi \; d^n \kappa \; d^n \lambda \; 
		\hat{\varphi}(\eta_\mu, \xi^\nu) \hat{\omega}(\kappa_\mu, \lambda^\nu) \nn \\
	& & \;\;\;\;\;\;\;\;\;\;\;\; 
		\times \; e^{- i(\eta_\mu X^\mu + \xi^\nu P_\nu)} \; 
		e^{- i(\kappa_\mu X^\mu + \lambda^\nu P_\nu)} \nn \\
	& = & \int d^n \eta \; d^n \xi \; d^n \kappa \; d^n \lambda \; 
		\hat{\varphi}(\eta_\mu, \xi^\nu) \hat{\omega}(\kappa_\mu, \lambda^\nu) \nn \\
	& & \;\;\;\;\;\;\;\;\;\;\;\; 
		\times e^{- i((\eta_\mu + \kappa_\mu)X^\mu + (\xi^\nu + \lambda^\nu)P_\nu) 
		- i \frac{\hbar}{2}\eta^{\mu\nu}( \eta_\mu \lambda_\nu - \xi_\nu \kappa_\mu ) \one }. \nn
\eeqarr
The final step we performed by the use of the Baker-Campbell-Hausdorff formula
\beq
	e^A \; e^B = e^{A + B + \frac{1}{2}[A,B] + \frac{1}{12}([A, [A, B]] - [B, [A, B]]) 
		+ \frac{1}{48}([A, [B, [B, A]]] - [B, [A, [A, B]]]) + \ldots)}. \nn
\eeq
We transform back by the use of the isomorphism $W$ and thus obtain
\beqarr
	(\varphi *_\hbar \omega)(x^\mu, p_\nu) 
		& = & \int d^n \eta \; d^n \xi \; d^n \kappa \; d^n \lambda \; 
			\hat{\varphi}(\eta_\mu, \xi^\nu) \hat{\omega}(\kappa_\mu, \lambda^\nu) \nn \\
		& & \;\;\;\;\;\;\;\;\;\;\;\; 
			\times e^{- i((\eta_\mu + \kappa_\mu)X^\mu + (\xi^\nu + \lambda^\nu)P_\nu )
			- i \frac{\hbar}{2}\eta^{\mu\nu}( \eta_\mu \lambda_\nu - \xi_\nu \kappa_\mu )} \nn\\
		& = & \int d^n \eta \; d^n \xi \; d^n \kappa \; d^n \lambda \; 
			\hat{\varphi}(\eta_\mu, \xi^\nu) \; e^{- i(\eta_\mu x^\mu + \xi^\nu p_\nu)} \nn \\
		& & \;\;\;\;\;\;\;\;\;\;\;\; 	
			\times \hat{\omega}(\kappa_\mu, \lambda^\nu) \; e^{- i(\kappa_\mu x^\mu + \lambda^\nu p_\nu)} 
			\; e^{- i \frac{\hbar}{2} \; \eta^{\mu\nu} ( \eta_\mu \lambda_\nu - \xi_\nu \kappa_\mu ) } \nn
\eeqarr
Replacing $\eta_\mu \to i \frac{\partial}{\partial x^\mu}, \xi_\nu \to i \frac{\partial}{\partial p^\nu}$ and $\kappa_\mu \to i \frac{\partial}{\partial \hat{q}^\mu}, \lambda_\nu \to i \frac{\partial}{\partial \hat{p}^\nu}$, 
we finally received the starproduct
\beqarr
	(\varphi *_\hbar \omega)(x^\mu, p_\nu) 
		& = & e^{+ i \frac{\hbar}{2} \; \eta^{\mu\nu}(\frac{\partial}{\partial x^\mu} 
			\frac{\partial}{\partial \hat{p}^\nu} - \frac{\partial}{\partial p^\nu} 
			\frac{\partial}{\partial \hat{x}_\mu} )} \; \varphi(x^\mu, p_\nu) \; \omega(\hat{x}^\mu, \hat{p}_\nu) 				
			|_{(\hat{x}^\mu, \hat{p}_\nu) \to (x^\mu, p_\nu)}. \nn \\
	\label{weylmoyalstarproduct}
\eeqarr
In particular for $\varphi(x^\rho, p_\sigma) = x^\rho$ and $\omega(x^\rho, p_\sigma) = p^\sigma$ we recover the 
second relation of (\ref{heisenbergideal}), distinguishing the generating relations of $U(\heisenberg_{2n})$ 
from those of $U(\Gamma)$.
\beqarr
	[x^\rho \stackrel{*_\hbar}{,} p^\sigma] & = & x^\rho *_\hbar p^\sigma - p^\sigma *_\hbar x^\rho \nn \\
	& = & e^{+ i \frac{\hbar}{2} \; \eta^{\mu\nu}(\frac{\partial}{\partial x^\mu} 
		\frac{\partial}{\partial \hat{p}^\nu} - \frac{\partial}{\partial p^\nu} 
		\frac{\partial}{\partial \hat{x}_\mu} )} x^\rho \cdot \hat{p}^\sigma |_{\hat{p}^\sigma \to p^\sigma} \nn \\
	& & \;\;\;\;\;\;\;\;\;\;\;\; - \; e^{+ i \frac{\hbar}{2} \; \eta^{\mu\nu}(\frac{\partial}{\partial x^\mu} 
		\frac{\partial}{\partial \hat{p}^\nu} - \frac{\partial}{\partial p^\nu} 
		\frac{\partial}{\partial \hat{x}_\mu} )} p^\sigma \cdot \hat{x}^\rho |_{\hat{x}^\rho \to x^\rho} \nn \\
	& = &  x^\rho \cdot p^\sigma + i \; \frac{\hbar}{2} \; \eta^{\rho\sigma}
		- p^\sigma \cdot x^\rho + i \; \frac{\hbar}{2} \; \eta^{\rho\sigma} \nn \\
	& = & i \; \eta^{\rho\sigma}. \nn
\eeqarr
This final computation closes our short review of Weyl-Moyal de\-for\-ma\-tion-quant\-iza\-tion. We are 
now prepared to formalise this procedure.

\section{Vector Fields $\mathfrak{W}(\Pi,\Gamma)$ on Minkowskian Phase Space}

Beginning with this section we formalise the presented constructions of Weyl and Moyal. In particular
we intend to absorb the starproduct (\ref{weylmoyalstarproduct}) into the modern setup of twists of
vector fields, as presented in \cite{Koch:2006cq}. We thus make a step beyond mere quantizations of Poisson 
manifolds because the twist formalism also enables us to further deform the Heisenberg-algebra 
$U(\heisenberg_{2n})$ itself. Furthermore the twist formalism also provides us with the opportunity 
to make required deformations of the Poincar\'e-algebra such that we can preserve spacetime 
covariance under deformations. In this section we therefore introduce the required Hopf-algebra of 
vector fields $\mathfrak{W}(\Pi,\Gamma)$ on $U(\Gamma)$ that provides us with the necessary tools to 
express the starproduct (\ref{weylmoyalstarproduct}) as a twist within $\mathfrak{W}(\Pi,\Gamma) \tensor \mathfrak{W}(\Pi,\Gamma)$. In the next section we accomodate $U(\mathfrak{p})$ within $\mathfrak{W}(\Pi,\Gamma)$ 
as a subalgebra. In this way, the starproduct turned into a twist thus also manages the deformation of $U(\mathfrak{p})$. In the mean time twists in $\mathfrak{W}(\Pi,\Gamma) \tensor \mathfrak{W}(\Pi,\Gamma)$ 
enable us, as already mentioned, to go beyond the quantization of Poisson manifolds. As already announced, 
double application of such twists then provides us with desired deformations of the Heisenberg-algebra $U(\heisenberg_{2n})$, covariant under corresponding deformations of $U(\mathfrak{p})$. In order to 
undertake this step of formalisation, we first consolidate our formulation of $U(\Gamma)$ by setting
\beqarr
	\xi^R & =	& \left\{
								\begin{array}{lcl}
									x^\rho & : & \rho = R \; \wedge \; R = 0, \ldots, (n-1) \\
									p^\mu  & : & \mu = R - n \; \wedge \; R = n, \ldots, (2n-1)
								\end{array}
							\right. \nn \\
	& & \nn \\
	& & \;\;\;\; R \in 0, \ldots, (n-1), n, \ldots, (2n-1).
	\label{xi}
\eeqarr
The generating relations (\ref{phasespaceideal}) of $U(\Gamma)$ are then reduced to the single equation
\beq
	\xi^R \xi^S - \xi^S \xi^R = 0.
	\label{xiphasespaceideal}
\eeq
As a first step towards the Hopf-algebra of vector fields $\mathfrak{W}(\Pi,\Gamma)$, we introduce 
an algebra of momenta $U(\Pi)$ in the following subsection.

\subsection{The Algebra of Momenta $U(\Pi)$ represented on $U(\Gamma)$}

In order to obtain a $2n$-dimensional Hopf-algebra of momenta $U(\Pi)$, we take a copy of 
$U(\Gamma)$ and enhance it to a Hopf-algebra. In praticular we consider $(\pi_N)_{N \in 0, \ldots 2n-1}$ 
as a basis for $U(\Pi)$. The generating relations, analogous to (\ref{phasespaceideal}), are then given by
$$
	\pi_M \cdot \pi_N - \pi_N \cdot \pi_M = 0, \;\;\;\; M, N \in 0, \ldots, (n-1), n, \ldots, (2n-1).
$$
The Hopf-structure on $U(\Pi)$ is given by the following coproduct, counit and antipode
$$
	\coproduct(\pi_M) = \pi_M \tensor \one_\pi + \one_\pi \tensor \pi_M, 
	\;\;\;\; \counit(\pi_M) = 0, 
	\;\;\;\; \antipode(\pi_M) = -\pi_M.
$$
The Hopf-algebra axioms are easily verified. The Hopf-algebra of momenta $U(\Pi)$ is represented 
by a left action on $U(\Gamma)$, as follows.
\beqarr
	\pi^M \laction \xi^R & = & - i E^{M R}, \nn \\
	\pi^M \laction \one & = & \counit(\pi^M), \nn \\
	\one \laction \xi^R & = & \xi^R, 
	\label{momentarepphasespace}
\eeqarr
To this purpose we introduce the $2n$-dimensional tensor 
$$ E^{M R} =	\left\{
										\begin{array}{lclcl}
											\eta^{M R} & : & M = 0, \ldots, (n-1) & \wedge & R = 0, \ldots, (n-1) \\
											0 & : & M = 0, \ldots, (n-1) & \wedge & R = n, \ldots, (2n-1) \\
											0 & : & M = n, \ldots, (2n-1) & \wedge & R = 0, \ldots, (n-1) \\ 
											\eta^{(M - n) (R - n)} & : & M = n, \ldots, (2n-1) & \wedge & R = n, \ldots, (2n-1)
										\end{array}
									\right.
$$
Alternatively we can also formulate (\ref{momentarepphasespace}) in the form
$$
	\pi_M \laction \xi^R = - i \Delta_M^R,
$$
with
$$ \Delta_M^R =	\left\{
										\begin{array}{lclcl}
											\delta_M^R & : & M = 0, \ldots, (n-1) & \wedge & R = 0, \ldots, (n-1) \\
											0 & : & M = 0, \ldots, (n-1) & \wedge & R = n, \ldots, (2n-1) \\
											0 & : & M = d, \ldots, (2n-1) & \wedge & R = 0, \ldots, (n-1) \\ 
											\delta_{M - n}^{R - n} & : & M = d, \ldots, (2n-1) & \wedge & R = d, \ldots, (2n-1)
										\end{array}
								\right. 
$$
We further verify that $U(\Pi)$ is realized on the vector space $\Gamma$ by
\beqarr
	(\pi_M \cdot \pi_N - \pi_N \cdot \pi_M) \laction \xi^R 
		& = & \pi_M \laction (\pi_N \laction \xi^R) - \pi_N \laction (\pi_M \laction \xi^R) \nn \\
		& = & - i \Delta_N^R \; \counit(\pi_M) + i \Delta_M^R \; \counit(\pi_N) = 0.
		\label{momentarealizedonphase}
\eeqarr
Moreover the action of $U(\Pi)$ respects the algebraic structure (\ref{xiphasespaceideal}) of $U(\Gamma)$, 
i.e. we have 
\beqarr
	\pi_M \laction ( \xi^R \cdot \xi^S - \xi^S \cdot \xi^R) & = & 
		\coproduct(\pi_M) \laction (\xi^R \cdot \xi^S) - \coproduct(\pi_M) \laction	(\xi^S \cdot \xi^R) \nn \\
	& = & (\pi_M \laction \xi^R) \xi^S + \xi^R (\pi_M \laction \xi^S) \nn \\
	& & \;\;\;\;\;\;\;\;\;\;\;\;\;\;\;\; 
	- (\pi_M \laction \xi^S) \xi^R - \xi^S (\pi_M \laction \xi^R) \nn \\
	& = & - i \Delta_M^R \; \xi^S - i \Delta_M^S \; \xi^R + i \Delta_M^S \; \xi^R + i \Delta_M^R \; \xi^S = 0. \nn \\
	\label{momentarespectphase}
\eeqarr
We thus obtained a valid representation of $U(\Pi)$ on $U(\Gamma)$ and can join them now to a single 
cross-product algebra.

\subsection{The Hopf-Algebra $\mathfrak{W}(\Pi,\Gamma)$ of Vector Fields}

In order to obtain the Hopf-algebra of vector fields $\mathfrak{W}(\Pi,\Gamma)$ on $U(\Gamma)$,
we have to consider the associative left cross-product algebra $U(\Gamma) \leftcrossproduct U(\Pi) \; $
that is build on the tensor product $U(\Gamma) \tensor U(\Pi)$. Additional division 
of this cross-product enables us to lift $\mathfrak{W}(\Pi,\Gamma)$ itself to a Hopf-algebra that is once 
more represented on $U(\Gamma)$. The left cross-product in $U(\Gamma) \tensor U(\Pi)$ is given by
\beqarr
	(\xi^R \tensor \pi^M) \odot (\xi^S \tensor \pi^N) & = & 
		\sum \xi^R (\pi^{M (1)} \laction \xi^S) \tensor \pi^{M (2)} \pi^N \nn \\
	& = & \xi^R (\pi^M \laction \xi^S) \tensor \pi^N
		+ \xi^R (\one \laction \xi^S) \tensor \pi^M \pi^N \nn \\
	& = & -i E^{M S}  (\xi^R \tensor \pi^N) + \xi^R \xi^S \tensor \pi^M \pi^N \nn \\
	& & \nn \\
	& & \coproduct(\pi^M) = \sum \pi^{M (1)} \tensor \pi^{M (2)}. 
	\label{phasespacecrossproduct}
\eeqarr
Within $U(\Gamma) \leftcrossproduct U(\Pi)$ the former subalgebras $U(\Gamma)$ and $U(\Pi)$ are also 
accomodated. They are identified by elements $\xi^R \equiv \xi^R \tensor \one$ and 
$\pi^M \equiv \one \tensor \pi^M$ respectively. We introduce the following elements
\beqarr
	\mathfrak{w}_0^{R M} := \xi^R \tensor \pi^M, & & \mathfrak{w}_+^M := \one \tensor \pi^M, \nn \\ 
	\mathfrak{w}_-^R := \xi^R \tensor \one, & & \one = \one \tensor \one, \label{minkowskivectorfields}
\eeqarr
that generate $U(\Gamma) \leftcrossproduct U(\Pi)$, i.e. we obtain
$$
	U(\Gamma) \leftcrossproduct U(\Pi) = \frac{T(U(\Gamma) \tensor U(\Pi))}{\mathcal{I}_{\Gamma, \Pi}},
$$
where $T(U(\Gamma) \tensor U(\Pi))$ is the tensor algebra of $U(\Gamma) \tensor U(\Pi)$ and 
$\mathcal{I}_{\Gamma, \Pi}$ is the two-sided ideal generated by relations
\beqarr
	\left[\mathfrak{w}_0^{R M}, \mathfrak{w}_0^{S N} \right]_\odot 
		= - i E^{M S} \mathfrak{w}_0^{R N} + i E^{N R} \mathfrak{w}_0^{S M}, & &
		\left[\mathfrak{w}_+^M, \mathfrak{w}_-^R \right]_\odot = - i E^{R M} \one \nn \\	
	\left[\mathfrak{w}_0^{R M}, \mathfrak{w}_-^S \right]_\odot = - i E^{S M} \mathfrak{w}_-^R, & &
		\left[\mathfrak{w}_0^{R M}, \mathfrak{w}_+^N \right]_\odot = i E^{R N} \mathfrak{w}_+^M, \nn \\
	\left[\mathfrak{w}_+^M, \mathfrak{w}_+^N \right]_\odot = 0, & &
		\left[\mathfrak{w}_-^R, \mathfrak{w}_-^S \right]_\odot = 0, 
	\label{minkowskicrossrelations}
\eeqarr
These are induced by (\ref{phasespacecrossproduct}) and (\ref{minkowskivectorfields}). We further enhance 
the ideal $\mathcal{I}_{\Gamma, \Pi}$ by setting $\mathfrak{w}_-^R = 0$ such that we receive a new 
two-sided ideal $\mathcal{I}_\mathfrak{W}$ that is generated by relations
\beqarr
	\left[\mathfrak{w}_0^{R M}, \mathfrak{w}_0^{S N} \right]_\odot 
		= - i E^{M S} \mathfrak{w}_0^{R N} + i E^{N R} \mathfrak{w}_0^{S M}, & &
		\left[\mathfrak{w}_0^{R M}, \mathfrak{w}_+^N \right]_\odot = i E^{R N} \mathfrak{w}_+^M, \nn \\	
	\left[\mathfrak{w}_+^M, \mathfrak{w}_+^N \right]_\odot = 0, & & 
	\label{minkowskivectorideal}
\eeqarr
such that we finally obtain the algebra of vector fields $\mathfrak{W}(\Pi,\Gamma)$ by
$$
	\mathfrak{W}(\Pi,\Gamma) = \frac{T(U(\Gamma) \tensor U(\Pi))}{\mathcal{I}_\mathfrak{W}}.
$$
The algebra $\mathfrak{W}(\Pi,\Gamma)$ is lifted to a Hopf-algebra by introducing coproducts, counits and 
antipodes on its generators $\mathfrak{w}_0^{R M}$ and $\mathfrak{w}_+^N$ according to
\beqarr
	& & \coproduct(\mathfrak{w}_0^{R M}) 
		= \mathfrak{w}_0^{R M} \tensor \one + \one \tensor \mathfrak{w}_0^{R M}, \;\;\;\;\;\;
		\counit(\mathfrak{w}_0^{R M}) = 0, \;\;\;\;\;\;
		\antipode(\mathfrak{w}_0^{R M}) = - \mathfrak{w}_0^{R M}, \nn \\
	& & \coproduct(\mathfrak{w}_+^M) 
			= \mathfrak{w}_+^M \tensor \one + \one \tensor \mathfrak{w}_+^M, \;\;\;\;\;\;
			\counit(\mathfrak{w}_+^M) = 0, \;\;\;\;\;\;
		\antipode(\mathfrak{w}_+^M) = - \mathfrak{w}_+^M. \nn
\eeqarr
It is easy to verify the axioms of Hopf-algebras and homomorphy. A detailed proof can be found in 
\cite{Koch:2006cq}. The Hopf-algebra of vector fields $\mathfrak{W}(\Pi,\Gamma)$ is represented by
a left action on $U(\Gamma)$ according to
\beqarr
	\mathfrak{w}_0^{R M} \laction \xi^S & = & -i E^{S M} \xi^R \nn \\
	\mathfrak{w}_+^M \laction \xi^R & = & -i E^{R M} \one \nn.
\eeqarr
We verify that the generating relations of $\mathfrak{W}(\Pi,\Gamma)$ are realized on $\Gamma$, i.e. 
for the first relation in  (\ref{minkowskivectorideal}) we obtain that
\beqarr
	& & (\mathfrak{w}_0^{R M} \cdot \mathfrak{w}_0^{S N} - \mathfrak{w}_0^{S N} \cdot \mathfrak{w}_0^{R M}
		+ i E^{M S} \mathfrak{w}_0^{R N} - i E^{N R} \mathfrak{w}_0^{S M})  \laction \xi^V \nn \\
	& & \;\;\;\;\;\;\;\; 
		= \mathfrak{w}_0^{R M} \laction (\mathfrak{w}_0^{S N} \laction \xi^V ) 
		- \mathfrak{w}_0^{S N} \laction (\mathfrak{w}_0^{R M} \laction \xi^V ) \nn \\
	& & \;\;\;\;\;\;\;\;\;\;\;\;\;\;\;\; 
		+ i E^{M S} (\mathfrak{w}_0^{R N}\laction \xi^V ) - i E^{N R} (\mathfrak{w}_0^{S M} \laction \xi^V) \nn \\
	& & \;\;\;\;\;\;\;\; 
		= - i E^{V N} (\mathfrak{w}_0^{R M} \laction \xi^S) 
		+ i E^{V M} (\mathfrak{w}_0^{S N} \laction \xi^R) \nn \\
	& & \;\;\;\;\;\;\;\;\;\;\;\;\;\;\;\; 
		+ E^{M S} E^{V N} \xi^R - E^{N R} E^{V M} \xi^S \nn \\
	& & \;\;\;\;\;\;\;\; 
		= - E^{V N} E^{S M} \xi^R
		+ E^{V M} E^{R N} \xi^S + E^{M S} E^{V N} \xi^R - E^{N R} E^{V M} \xi^S = 0. \nn
\eeqarr
For the second relation we further compute that
\beqarr
	& & (\mathfrak{w}_0^{R M} \cdot \mathfrak{w}_+^N - \mathfrak{w}_+^N \cdot \mathfrak{w}_0^{R M} 
		- i E^{R N} \mathfrak{w}_+^M) \laction \xi^V \nn \\
	& & \;\;\;\;\;\;\;\; = \mathfrak{w}_0^{R M} \laction (\mathfrak{w}_+^N \laction \xi^V)
		- \mathfrak{w}_+^N \laction (\mathfrak{w}_0^{R M} \laction \xi^V)
		- i E^{R N} (\mathfrak{w}_+^M \laction \xi^V) \nn \\
	& & \;\;\;\;\;\;\;\; = - i E^{V N} (\mathfrak{w}_0^{R M} \laction \one)
		+ i E^{V M} (\mathfrak{w}_+^N \laction \xi^R)
		- E^{R N} E^{V M}  \nn \\
	& & \;\;\;\;\;\;\;\; = E^{V M} E^{R N} - E^{R N} E^{V M} = 0. \nn
\eeqarr		
The third relation is already satisfied by (\ref{momentarealizedonphase}). We further check that
$\mathfrak{W}(\Pi,\Gamma)$ respects the generating relations of $U(\Gamma)$. For $\mathfrak{w}_+^M$
this is already verified by (\ref{momentarespectphase}). We thus consider
\beqarr
	& & \mathfrak{w}_0^{R M} \laction (\xi^V \xi^W - \xi^W \xi^V) \nn \\
	& & \;\;\;\;\;\;\;\; = \coproduct(\mathfrak{w}_0^{R M}) \laction (\xi^V \xi^W)
		- \coproduct(\mathfrak{w}_0^{R M}) \laction (\xi^W \xi^V) \nn \\
	& & \;\;\;\;\;\;\;\; = (\mathfrak{w}_0^{R M} \laction \xi^V) \xi^W 
		+ \xi^V (\mathfrak{w}_0^{R M} \laction \xi^W) \nn \\
	& & \;\;\;\;\;\;\;\;\;\;\;\;\;\;\;\; 	
		- (\mathfrak{w}_0^{R M} \laction \xi^W) \xi^V
		- \xi^W (\mathfrak{w}_0^{R M} \laction \xi^V) \nn \\
	& & \;\;\;\;\;\;\;\; = - i E^{V M} \xi^R \xi^W 
		- i E^{W M} \xi^V \xi^R + i E^{W M} \xi^R \xi^V
		+  i E^{V M} \xi^W \xi^R = 0. \nn 	
\eeqarr

We are now prepared to take the next step that embeds the Poincar\'e-algebra $U(\mathfrak{p})$
within $\mathfrak{W}(\Pi,\Gamma)$.

\section{The Vector Field Representation of the Lorentz-Algebra}
 
The previous preparations of the last sections enable us to represent the Poincar\'e-algebra 
$U(\mathfrak{p})$ within $\mathfrak{W}(\Pi,\Gamma)$. As a corresponding representation of 
the Lorentz-generators $M^{LN} \in U(so_{1,n-1})$ we introduce
\beqarr
	M^{L N} & =	& \left\{
										\begin{array}{lclcl}
											\mathfrak{w}_0^{L N} - \mathfrak{w}_0^{N L} 
												& : & L = 0, \ldots, (n-1) & \wedge & N = 0, \ldots, (n-1) \\
											0 & : & L = 0, \ldots, (n-1) & \wedge & N = n, \ldots, (2n-1) \\
											0 & : & L = n, \ldots, (2n-1) & \wedge & N = 0, \ldots, (n-1) \\ 
											\mathfrak{w}_0^{L N} - \mathfrak{w}_0^{N L} 
												& : & l = n, \ldots, (2n-1) & \wedge & N = n, \ldots, (2n-1)
										\end{array}
									\right. \nn \\
	& & \nn \\
	\label{lorentzgeneratorphase}
\eeqarr
Translational operators are already given by the algebra of momenta $U(\Pi)$, i.e. we have
\beq
	P^N := \mathfrak{w}_+^N.
	\label{translationgeneratorphase}
\eeq
With relations (\ref{minkowskivectorideal}) we compute the generating relations (\ref{poincarerelations}) 
of $U(\mathfrak{p})$ in their block-diagonal form (\ref{blockdiagonal}) to be
\beqarr
	\left[ M^{L N}, M^{I P} \right] & = & - i E^{N I} M^{L P} + i E^{P L} M^{I N} 
		+ i E^{N P} M^{L I} - i E^{I L} M^{P N} \nn \\
	\left[ M^{L N}, P^M \right] & = & i E^{L M} P^N - i E^{N M} P^L \nn
\eeqarr
Due to the linaer combination of the Lorentz generators (\ref{lorentzgeneratorphase}) in terms of 
generators of $\mathfrak{W}(\Pi,\Gamma)$, the Hopf-structure of the vector fields is carried over to the
expected Hopf-structure in this representation of $U(\mathfrak{p})$, i.e. we have
$$
	\coproduct(M^{L N}) = M^{L N} \tensor \one + \one \tensor M^{L N},
	\;\;\;\; \counit(M^{L N}) = 0,
	\;\;\;\; \antipode(M^{L N}) = - M^{L N}.
$$
The representation of $\mathfrak{W}(\Pi,\Gamma)$ on $U(\Gamma)$ determines that of $U(\mathfrak{p})$, 
i.e.
$$
	M^{L N} \laction \xi^R = i E^{N R} \xi^L - i E^{L R} \xi^N.
$$
According to (\ref{xi}), we receive the corresponding $n+n$-decomposition being
\beqarr
	m^{\mu\nu} \laction x^\rho & = & - i \eta^{\nu \rho} x^\mu + i \eta^{\mu \rho} x^\nu,  \nn \\
	m^{\mu\nu} \laction p^\sigma & = & - i \eta^{\nu \rho} p^\mu + i \eta^{\mu \rho} p^\nu, \nn
\eeqarr
that is in accordance with (\ref{reppoincareminkowski}). Since $U(\mathfrak{p})$ is a sub-Hopf-algebra 
of $\mathfrak{W}(\Pi,\Gamma)$, we do not require to further verify properties of the representation 
of $U(\mathfrak{p})$ on $U(\Gamma)$. Before we turn to actual twist-deformations of $U(\Gamma)$ 
and $U(\heisenberg_{2n})$, we have to consider the twist formalism as such. In particular we 
have to discuss now double application of twists.

\section{Twisting}

In this section we first shortly review basic definitions and properties of twists. Our primary 
aim however is to show that a double application of twists in turn can be treated as a twist as well.
This comes in handy when we first deform the $2n$-dimensional commutative phase space algebra 
$U(\Gamma)$ to the Heisenberg-algebra $U(\heisenberg_{2n})$ and in a second step twist once more
in order to obtain some deformation of $U(\heisenberg_{2n})$ itself. These two twists of course 
can be merged to a single expression by the use of the Baker-Campbell-Hausdorff formula. But 
application of this formula might turn out to be complicated by the computation of higher order 
terms in the exponent. It might thus be a better choice not to evaluate this product 
of twists, althought the application then becomes a little bulky. Thus, up to the double application 
of twists, the first subsection of this section is rather a review to keep everything clear.
The second subsection further embeds the starproduct of Weyl and Moyal (\ref{weylmoyalstarproduct}) into
the vector field formalism. 

\subsection{Double Twisting}

As announced, we begin with a little review of the definition of twists and basic properties.
We thus define twists for a general Hopf-algebra $\hopfalgebra$ to be given by

\begin{definition} Let $\left(\hopfalgebra, \product, \unit, \coproduct, \counit, \antipode ; 
\field \right)$ be a Hopf-algebra over the field $\field$. Then an invertible object 
$\twist \in \hopfalgebra \tensor \hopfalgebra$ is called a \emph{twist}, if the following 
two conditions hold
\beqarr
	\twist_{1 2}\left(\coproduct \tensor \id \right)(\twist) 
		& = & \twist_{2 3} \left(\id \tensor \coproduct \right)(\twist) \label{firstdeftwist} \\
	\left(\counit \tensor \id \right)(\twist) & = 1 = & \left( \id \tensor \counit \right)(\twist). 
		\label{secdeftwist} 
\eeqarr
For $\twist = \sum \twist^{(1)} \tensor \twist^{(2)}$ the objects $\twist_{1 2}$ and $\twist_{2 3}$ 
are defined by
\beqarr
	\twist_{1 2} & = & \sum \twist^{(1)} \tensor \twist^{(2)} \tensor \one \nn \\
	\twist_{2 3} & = & \sum \one \tensor \twist^{(1)} \tensor \twist^{(2)}. \nn
\eeqarr
\end{definition}

This definition is the basic ingredient to perform deformations. That these twists in turn provide the 
desired deformations of Hopf-algebras $\hopfalgebra_\twist$ is stated within the following proposition.

\begin{proposition} Let $\left(\hopfalgebra, \product, \unit, \coproduct, \counit, \antipode; 
	\field \right)$ be a Hopf-algebra and let furthermore the objects $\eta, \eta^{-1} \in \hopfalgebra$ 
	be given by 
	\beqarr
		\eta & = & \product \left(\id \tensor \antipode \right)(\twist) \nn \\
		\eta^{-1} & = & \product \left(\antipode \tensor \id \right) (\twist).\nn 
	\eeqarr
	Then $\left(\hopfalgebra, \product, \unit, \coproduct_\twist, \counit, \antipode_\twist; \field \right)$
	with 
	\beqarr
		\coproduct_\twist (h) & = & \twist \coproduct (h) \twist^{-1} \nn \\
		\antipode_\twist (h) & = & \eta \antipode(h)\eta^{-1}\nn
	\eeqarr
	and $h\in \hopfalgebra$ is the Hopf-algebra $\hopfalgebra_\twist$ that is called the 
	\emph{twist of} $\hopfalgebra$. 
\end{proposition}

The crucial point we have to emphasize within the next step is that the Hopf-algebra 
$\hopfalgebra$ is not necessarily cocommutative. And in this respect $\hopfalgebra$
might already be the outcome of a preceding twist, applied to a Hopf-algebra that actually
might have been cocommutative. Lets thus assume that we have a twist
$\mathcal{J} \in \hopfalgebra \tensor \hopfalgebra$ in the tensor product of a Hopf-algebra 
$\hopfalgebra$. In particular it satisfies conditions (\ref{firstdeftwist}) and 
(\ref{secdeftwist}) of above definition, i.e. 
\beqarr
	\mathcal{J}_{1 2}\left(\coproduct \tensor \id \right)(\mathcal{J}) 
		& = & \mathcal{J}_{2 3} \left(\id \tensor \coproduct \right)(\mathcal{J}) \label{jtwistrel} \\
	\left(\counit \tensor \id \right)(\mathcal{J}) & = 1 = & \left( \id \tensor \counit \right)(\mathcal{J}). \nn 
\eeqarr
We then receive a Hopf-algebra $\hopfalgebra_\mathcal{J}$ according to above proposition. We can 
now go ahead and twist once more. Thus let $\mathcal{G} \in \hopfalgebra \tensor \hopfalgebra$
be a twist of $\hopfalgebra_\mathcal{J}$, i.e. we have 
\beqarr
	\mathcal{G}_{1 2}\left(\coproduct_{\mathcal{J}} \tensor \id \right)(\mathcal{G}) 
		& = & \mathcal{G}_{2 3} \left(\id \tensor \coproduct_\mathcal{J} \right)(\mathcal{G}) \nn \\
	\left(\counit \tensor \id \right)(\mathcal{G}) & = 1 = & \left( \id \tensor \counit \right)(\mathcal{G}). \nn 
\eeqarr
With $\coproduct_\mathcal{J}(h) = \mathcal{J} \coproduct(h) \mathcal{J}^{-1}$ for $h \in \hopfalgebra$ 
the first of these two relations can be writen as
\beq
	\mathcal{G}_{1 2} \mathcal{J}_{1 2} \left(\coproduct \tensor \id \right)(\mathcal{G}) \mathcal{J}^{-1}_{1 2}
		= \mathcal{G}_{2 3} \mathcal{J}_{2 3}\left(\id \tensor \coproduct \right)(\mathcal{G}) \mathcal{J}^{-1}_{2 3}.
	\label{gtwistrel}
\eeq
We thus claim that $\twist = \mathcal{G} \cdot \mathcal{J}$ is a twist of $\hopfalgebra$ as well. 
Relation (\ref{secdeftwist}) is directly satisfied by the homomorphy property of the counit $\counit$. 
Relation (\ref{firstdeftwist}) in turn is verified by direct computation. In particular we obtain by 
the use of (\ref{gtwistrel}) and (\ref{jtwistrel}) that
\beqarr
	\twist_{1 2} \left( \coproduct \tensor \id \right)(\twist)
	& = & \mathcal{G}_{1 2} \cdot \mathcal{J}_{1 2} 
		\left( \coproduct \tensor \id \right)(\mathcal{G} \cdot \mathcal{J}) \nn \\
	& = & \mathcal{G}_{1 2} \cdot \mathcal{J}_{1 2} 
		\left( \coproduct \tensor \id \right)(\mathcal{G}) 
		\left( \coproduct \tensor \id \right)(\mathcal{J}) \nn \\	
	& = & \mathcal{G}_{1 2} \cdot \mathcal{J}_{1 2} 
		\left( \coproduct \tensor \id \right)(\mathcal{G}) 
		\mathcal{J}^{-1}_{1 2} \mathcal{J}_{2 3} \left( \id \tensor \coproduct \right)(\mathcal{J}) \nn \\
	& = & \mathcal{G}_{2 3} \mathcal{J}_{2 3}
		\left(\id \tensor \coproduct \right)(\mathcal{G}) 
		\mathcal{J}^{-1}_{2 3} \mathcal{J}_{2 3} \left( \id \tensor \coproduct \right)(\mathcal{J}) \nn \\
	& = & \mathcal{G}_{2 3} \mathcal{J}_{2 3}
		\left(\id \tensor \coproduct \right)(\mathcal{G}) \left( \id \tensor \coproduct \right)(\mathcal{J}) \nn \\
	& = & \twist_{2 3} \left(\id \tensor \coproduct \right)(\twist). \nn
\eeqarr
We thus collected all required ingredients to proceed to actual deformations of $U(\heisenberg_{2n})$.

\subsection{Twists, Starproducts and Vector Fields}

The Hopf-algebra of vector fields $\mathfrak{W}(\Pi,\Gamma)$ enables us to express the starproduct (\ref{weylmoyalstarproduct}) as the inverse of a twist $\mathcal{G} \in \mathfrak{W}(\Pi,\Gamma) 
\tensor \mathfrak{W}(\Pi,\Gamma)$ that in the mean time is capable to deform the Poincar\'e-algebra
$U(\mathfrak{p})$. The twist $\mathcal{G}$ corresponding to starproduct (\ref{weylmoyalstarproduct}) 
is given by
\beq
	\mathcal{G} = e^{+ i \frac{\hbar}{2} \; \Xi_{M N} \; \mathfrak{w}_+^M \tensor \mathfrak{w}_+^N}, \;\;\;\;
	\label{itwist}
\eeq
where we define the antisymmetric tensor $\Xi_{M N}$ by 
$$ \Xi_{M N} =	\left\{
										\begin{array}{lclcl}
											0 & : & M = 0, \ldots, (n-1) & \wedge & N = 0, \ldots, (n-1) \\
											\eta_{M, (N - n)} & : & M = 0, \ldots, (n-1) & \wedge & N = n, \ldots, (2n-1) \\
											- \eta_{(M - n), N} & : & M = n, \ldots, (2n-1) & \wedge & N = 0, \ldots, (n-1) \\ 
											0 & : & M = n, \ldots, (2n-1) & \wedge & N = n, \ldots, (2n-1)
										\end{array}
									\right.
$$
The defining conditions (\ref{firstdeftwist}) and (\ref{secdeftwist}) for twists are easily checked. It 
is also easily verified that the generating relations (\ref{heisenbergideal}) of $U(\heisenberg_{2n})$ 
are reproduced by the inverse $\mathcal{G}^{-1}$ of (\ref{itwist}). We can thus use (\ref{itwist}) in 
order to deform the algebraic sector of $U(\Gamma)$ to that of $U(\heisenberg_{2n})$. We further 
concentrate on the deformation of coproducts (\ref{coproductpoincare}) in $U(\mathfrak{p})$ within the 
representation (\ref{lorentzgeneratorphase}). Due to commutativity between $P^M$ and $\mathcal{G}$ we
only expect possible deformations for the coproduct of $M^{L N}$. With the undeformed coproduct
$$
	\coproduct(M^{L N}) = \coproduct(\mathfrak{w}_0^{L N} - \mathfrak{w}_0^{N L})
		= (\mathfrak{w}_0^{L N} - \mathfrak{w}_0^{N L}) \tensor \one + 
		\one \tensor (\mathfrak{w}_0^{L N} - \mathfrak{w}_0^{N L})
$$
and with the help of the formula
$$
	e^A \; B \; e^{-A} = \sum_{n = 0}^\infty \frac{1}{n !}
		\left[A, \left[A, \left[A, \ldots \left[A, B \right] \right] \right] \right],
$$
we compute the deformed coproduct to be 
\beqarr
	\coproduct(M^{L N}) & = & M^{L N} \tensor \one + \one \tensor M^{L N} \nn \\
	& & \;\;\;\;\;\;\;\; + \frac{\hbar}{2} \; \Xi_{R S} \; 
		(E^{R L} \mathfrak{w}_+^N  - E^{R N} \mathfrak{w}_+^L ) \tensor \mathfrak{w}_+^S \nn \\
	& & \;\;\;\;\;\;\;\; + \frac{\hbar}{2} \; \Xi_{R S} \; \mathfrak{w}_+^R  \tensor 
		(E^{S L} \mathfrak{w}_+^N  - E^{S N} \mathfrak{w}_+^L ) \nn
\eeqarr 
This corresponds to results presented in \cite{Chaichian:2004za,Koch:2004ud,Oeckl:2000eg,Wess:2003da}. However, 
we should give some comments to this particular deformed coproduct in respect to the discussion of the introduction. 
Textbooks on field theory of course never mention the existence of a deformed coproduct of the Poincar\'e-algebra $U(\mathfrak{p})$ in order to respect the commutation relations of the Heisenberg-algebra $U(\heisenberg_{2n})$. Within the introduction we argued that $U(\mathfrak{p})$ could be embedded in $U(\heisenberg_{2n})$ - without 
the requirement to explicitly introduce any deformed coproducts. In fact the coproducts of $U(\mathfrak{p})$ 
actually are deformed without being manifest. This can be seen as follows. We freely choose the upper part of the block-diagonal generator $M^{L N}$ and consider its coproduct, i.e.
\beqarr
	\coproduct(M^{\lambda \nu}) & = & M^{\lambda \nu} \tensor \one + \one \tensor M^{\lambda \nu}
		+ \frac{\hbar}{2} \; \Xi_{R S} \; 
		(E^{R \lambda} \mathfrak{w}_+^\nu  - E^{R \nu} \mathfrak{w}_+^\lambda ) \tensor \mathfrak{w}_+^S \nn \\
	& & \;\;\;\;\;\;\;\; + \frac{\hbar}{2} \; \Xi_{R S} \; \mathfrak{w}_+^R  \tensor 
		(E^{S \lambda} \mathfrak{w}_+^\nu  - E^{S \nu} \mathfrak{w}_+^\lambda ) \nn \\
	& = & M^{\lambda \nu} \tensor \one + \one \tensor M^{\lambda \nu}
		+ \frac{\hbar}{2} \; \eta_{\rho \sigma} \; 
		(\eta^{\rho \lambda} \mathfrak{w}_+^\nu  
		- \eta^{\rho \nu} \mathfrak{w}_+^\lambda ) \tensor \mathfrak{w}_+^{(\sigma + n)} \nn \\
	& & \;\;\;\;\;\;\;\; - \frac{\hbar}{2} \; \eta _{\rho \sigma} \; \mathfrak{w}_+^{(\sigma + n)} \tensor 
		(\eta^{\rho \lambda} \mathfrak{w}_+^\nu - \eta^{\rho \nu} \mathfrak{w}_+^\lambda ) \nn
\eeqarr 
We see that the coproduct of $M^{\lambda \nu}$ is nearly cocommutative - up to a minus sign in the deformed part.
A cocommutative deformation would be trivial and thus we have a true but hidden deformation for the case we 
embed $U(\mathfrak{p})$ in $U(\heisenberg_{2n})$ as we did in the introduction. This particular minus sign distinguishes the naive "action" of the momentum on a coordinate $[p^\mu, x^\rho]$ from the "action" of a 
coordinate on momentum $[x^\mu, p^\rho]$. This comes into account when we determine the representation of 
$m^{\mu\nu} = x^\mu p^\nu - x^\nu p^\mu$ on a coordinate or a momentum operator by the commutator $[\;, \;]$ as
in (\ref{xmaction}) and (\ref{pmaction}).

\section{An Example for a Twisted Heisenberg-Algebra}

In this final section we intend to outline the presented construction for a specific example. In particular 
we concentrate on how the Heisenberg-algebra $U(\heisenberg_{2n})$ is further deformed by a an additional twist 
$\mathcal{I}$. This corresponds to a second deformation of $U(\Gamma)$. In this section we merely whish to give some guidance to the presented apperatus and thus stick to a very simple but nontrivial example. We leave it to 
the reader to find more interesting or even more realistic deformations. We sketch an example that corresponds 
to a twist presented earlier in \cite{Koch:2006cq} and \cite{Jambor:2004kc} and adapt it to our context. 
This specific twist is given by
\beq
	\mathcal{I} = e^{i \; a \; \mathfrak{w}_0^{(2n-1) (2n-1)} \; \tensor \; \mathfrak{w}_+^{(n - 1)}},
	\;\;\;\; a \in \R. \nn
\eeq
With the total twist
$$
	\twist = \mathcal{G} \cdot \mathcal{I} = 
	e^{+ i \frac{\hbar}{2} \; \Xi_{M N} \; \mathfrak{w}_+^M \tensor \mathfrak{w}_+^N} \cdot
	e^{+ i \; a \; \mathfrak{w}_0^{(2n-1) (2n-1)} \; \tensor \; \mathfrak{w}_+^{(n - 1)}},
$$
we obtain the starproduct 
$$
	\twist^{-1} = \mathcal{I}^{-1} \cdot \mathcal{G}^{-1} = 
	e^{- i \; a \; \mathfrak{w}_0^{(2n-1) (2n-1)} \; \tensor \; \mathfrak{w}_+^{(n - 1)}} \cdot
	e^{- i \frac{\hbar}{2} \; \Xi_{M N} \; \mathfrak{w}_+^M \tensor \mathfrak{w}_+^N},
$$
that provides us with a deformation of $U(\heisenberg_{2n})$. With the starproduct $\mathcal{I}^{-1}$
only some of the generating relations of $U(\heisenberg_{2n})$ actually become deformed. We first 
generally consider the starproduct of the product of two generators $\xi^R, \xi^S \in U(\Gamma)$, i.e.
$$
	\xi^R *_\twist \xi^S = \xi^R \xi^S + i \; a \; E^{(2n-1) R} E^{(n-1)S} \xi^{(2n-1)} 
	+ i \; \frac{\hbar}{2} \; \Xi^{R S}.
$$
In particular we thus obtain for the choice $R \to 2n-1$ and $S \to n-1$ that
$$
	\xi^{(2n-1)} *_\twist \xi^{(n-1)} = \xi^{(2n-1)} \xi^{(n-1)}  + i \; a \; \xi^{(2n-1)} 
	- i \frac{\hbar}{2} \eta^{(n-1)(n-1)},
$$
such that within the $n+n$-separation we obtain the commutator
$$
	\left[x^{(n-1)} \stackrel{*_\twist}{,} p^{(n-1)} \right] = i \; \hbar \; \eta^{(n-1)(n-1)} - i \; a \; p^{(n-1)}.
$$
We thus obtained the expected deformation of the Heisenberg-algebra $U(\heisenberg_{2n})$ for one of its
most characteristic relations. We further compute an example for a deformation of the coproduct of $M^{L N}$ such 
that we obtain manifest covariance with respect to deformed $U(\mathfrak{p})$. In particular we choose the 
coproduct $\coproduct_\twist(M^{(2n-1) n})$ and to this purpose we first compute the corresponding twisted 
coproducts of $\mathfrak{w}_0^{(2n-1) n}$ and $\mathfrak{w}_0^{n (2n-1)}$, i.e. we have
\beqarr
	\coproduct_\twist(\mathfrak{w}_0^{(2n-1) n}) & = & \mathcal{G} \cdot \mathcal{I}
		\left(\mathfrak{w}_0^{(2n-1) n} \tensor \one + \one \tensor \mathfrak{w}_0^{(2n-1) n} \right) 
		\mathcal{I}^{-1} \cdot \mathcal{G}^{-1} \nn \\
		& = & \mathcal{G}
		\left(\mathfrak{w}_0^{(2n-1) n} \tensor \one + \one \tensor \mathfrak{w}_0^{(2n-1) n} \right. \nn \\
		& & \left. + \mathfrak{w}_0^{(2n-1) n} 
		\tensor (e^{+ a \; \eta^{(n-1)(n-1)} \; \mathfrak{w}_+^{(n-1)}} - 1)\right) 
		\mathcal{G}^{-1} \nn \\
		& = & \mathfrak{w}_0^{(2n-1) n} \tensor e^{+ a \; \eta^{(n-1)(n-1)} \; \mathfrak{w}_+^{(n-1)}} 
		+ \one \tensor \mathfrak{w}_0^{(2n-1) n} \nn \\
		& & - \frac{\hbar}{2} \; \eta^{(n-1)(n-1)} \; \mathfrak{w}_+^n 
			\tensor \mathfrak{w}_+^{(2n-1)} e^{+ a \; \eta^{(n-1)(n-1)} \; \mathfrak{w}_+^{(n-1)}}  \nn \\
		& &	+ \frac{\hbar}{2} \; \eta^{(n-1)(n-1)} \; \mathfrak{w}_+^{(2n-1)} \tensor \mathfrak{w}_+^n \nn
\eeqarr
and
\beqarr
	\coproduct_\twist(\mathfrak{w}_0^{n (2n-1)}) & = & \mathcal{G} \cdot \mathcal{I}
		\left(\mathfrak{w}_0^{n (2n-1)} \tensor \one + \one \tensor \mathfrak{w}_0^{n (2n-1)} \right) 
		\mathcal{I}^{-1} \cdot \mathcal{G}^{-1} \nn \\
		& = & \mathcal{G}
		\left(\mathfrak{w}_0^{n (2n-1)} \tensor \one + \one \tensor \mathfrak{w}_0^{n (2n-1)} \right. \nn \\
		& & \left. + \mathfrak{w}_0^{n (2n-1)} 
		\tensor (e^{- a \; \eta^{(n-1)(n-1)} \; \mathfrak{w}_+^{(n-1)}} - 1) \right) 
		\mathcal{G}^{-1} \nn \\
		& = & \mathfrak{w}_0^{n (2n-1)} \tensor e^{- a \; \eta^{(n-1)(n-1)} \; \mathfrak{w}_+^{(n-1)}} 
		+ \one \tensor \mathfrak{w}_0^{n (2n-1)} \nn \\
		& & + \frac{\hbar}{2} \; \eta^{(n-1)(n-1)} \; \mathfrak{w}_+^{(2n-1)} 
		\tensor \mathfrak{w}_+^n e^{- a \; \eta^{(n-1)(n-1)} \; \mathfrak{w}_+^{(n-1)}} \nn \\
		& & - \frac{\hbar}{2} \; \eta^{(n-1)(n-1)} \; \mathfrak{w}_+^n \tensor \mathfrak{w}_+^{(2n-1)}. \nn
\eeqarr
We thus obtain that 
\beqarr
	& & \coproduct_\twist(M^{(2n-1) n}) \nn \\
	& & \;\;\;\;\;\;\;\; = M^{(2n-1) n} \tensor e^{+ a \; \eta^{(n-1)(n-1)} \; P^{(n-1)}}
	+ \one \tensor M^{(2n-1) n} \nn \\
	& & \;\;\;\;\;\;\;\; + 2 \; \mathfrak{w}_0^{n (2n-1)} \tensor \sinh (+ a \; \eta^{(n-1)(n-1)} \; P^{(n-1)}) \nn \\
	& & \;\;\;\;\;\;\;\; - \frac{\hbar}{2} \; \eta^{(n-1)(n-1)} \; P^n 
			\tensor \left( P^{(2n-1)} e^{+ a \; \eta^{(n-1)(n-1)} \; P^{(n-1)}}  
			- P^{(2n-1)} \right) \nn \\
	& &	\;\;\;\;\;\;\;\; + \frac{\hbar}{2} \; \eta^{(n-1)(n-1)} \; P^{(2n-1)} \tensor 
		\left( P^n - P^n e^{- a \; \eta^{(n-1)(n-1)} \; P^{(n-1)}} \right) \nn 
\eeqarr
There are of course several more deformed coproducts for this specific example of deformation.
However, since we merely whish to give some idea of how the constructions in this work are 
applied to specific examples, we close our considerations at this point. 

\section{Conclusion}

Providing the formalism to perform deformations of the Heisenberg-algebra and the corresponding twists
of the Poincar\'e-algebra is certainly only one step of several that have to be mastered in order to 
obtain some enhanced version of relativistic quantum mechanics. In order to receive useful representations 
of the deformed Heisenberg-algebra on states of a Hilbert-space, it is for example a crucial point to discuss
hermiticity and self-adjointness of the generators in deformed $U(\heisenberg_{2n})$. It is moreover not yet
clear what further implications for the interpretation of quantum mechanics might result form such algebraic 
mixture of coordinates and momenta. However, quantum mechanics as we apply it in field theories, does not
discribe high-energy measurements in such a way as we would expect them from Planck-scale physics. Thus, 
regarding noncommutative geometry as a high-energy approach, we should als take into account that gravity 
might not only provide a static form of noncommutativity - but one that is caused by the properties of 
matter itself that exists within such backgrounds.

\section{Acknowledgment}

We thank Prof. Dr. Julius Wess for his support.

\end{document}